\documentclass[prb,final,showpacs,twocolumn,superscriptaddress,aps]{revtex4-2}

\usepackage[utf8]{inputenc}
\usepackage[american,british]{babel}
\usepackage[T1]{fontenc}
\usepackage[pdftex]{graphicx}  
\usepackage{graphicx, xcolor}
\usepackage{dcolumn}
\usepackage{braket}
\usepackage{bm}
\usepackage{amsmath,amsthm,amssymb}
\usepackage{color}
\usepackage{verbatim}
\usepackage{lipsum}

\usepackage[colorlinks,linkcolor=blue,citecolor=blue,urlcolor=blue]{hyperref}

\newcommand{\me}[1]{\left\langle #1 \right\rangle}
\newcommand{\TraceBracket}[1]{\mathrm{Tr}\Big( #1 \Big)}
\newcommand{\ee}{\mathrm{e}}
\newcommand{\Pauli}{\hat{\sigma}}
\newcommand{\Ham}{\hat{H}}
\newcommand{\period}{\tau}
  
\usepackage{dsfont}
\usepackage{physics}
\usepackage{soul}
\newcommand{\revision}[1]{{\textcolor{black}{#1}}}

\begin{document}
\title{Clean two-dimensional Floquet time-crystal}
\author{Alessandro Santini}
\email{asantini@sissa.it}
\affiliation{SISSA, via Bonomea 265, 34136 Trieste, Italy}

\author{Giuseppe E. Santoro}
\email{santoro@sissa.it}
\affiliation{SISSA, via Bonomea 265, 34136 Trieste, Italy}
\affiliation{International Centre for Theoretical Physics (ICTP), P.O. Box 586, I-34014 Trieste, Italy}
\affiliation{CNR-IOM Democritos National Simulation Center, Via Bonomea 265, I-34136 Trieste, Italy}

\author{Mario Collura}
\email{mcollura@sissa.it}
\affiliation{SISSA, via Bonomea 265, 34136 Trieste, Italy}
\affiliation{INFN Sezione di Trieste, via Bonomea 265, 34136 Trieste, Italy}
\date{\today}

\begin{abstract}
We consider the two-dimensional quantum Ising model, in absence of disorder, subject to periodic imperfect global spin flips. We show by a combination of exact diagonalization and tensor-network methods that the system can sustain a spontaneously broken discrete time-translation symmetry. Employing careful scaling analysis, we show the feasibility of a two-dimensional discrete time-crystal (DTC) \textcolor{black}{pre-thermal} phase. Despite an unbounded energy pumped into the system, in the high-frequency limit, a well-defined effective Hamiltonian controls a finite-temperature intermediate regime, wherein local time averages are described by thermal averages.
As a consequence, the long-lived stability of the DTC relies on the existence of a long-range ordeblack phase at finite temperature.
Interestingly, even for large deviations 
from the perfect spin flip, we observe a non-perturbative change in the decay rate of the order parameter, which is related to the long-lived stability of the magnetic domains in 2D.
\end{abstract}

\keywords{Discrete time-crystals; Two-dimensional quantum physics; TDVP; Many-body quantum dynamics} 
 
\maketitle

\section{Introduction}
The fast advancements of quantum technologies and the exceptional progress and precision achieved in the experimental realizations of closed quantum systems made quantum non-equilibrium physics a new exciting field. 
A decade ago, Wilczek proposed a fascinating idea for a novel phase of matter, dubbed ``time crystal'', which involves the spontaneous breaking of the continuous time-translational symmetry of the system \cite{WilczekPRL12}. While a number of no-go theorems ruled out the existence of such a phase of matter \cite{BrunoPRL13,WatanabeOshikawaPRL15}, the possibility of breaking a ``discrete'' time-translation symmetry --- realising discrete time crystals (DTC) --- has generated a large body of literature ~\cite{Sacha17review,khemani2019brief,ElseReview20,ElsePRL16,ElsePRX17_PrethermalDTC,KhemaniPRL16,VonKeyserlingkPRB16_AbsoluteStability,zhang2017observation,RussomannoPRB17_LMGDTC,PhysRevB.103.014305,PhysRevB.103.224311,RussomannoPRR20,PizziPRB20_CleanDTC,Hahn_2021,collura2021discrete,giachetti2022high,PhysRevResearch.4.023018}. 

In a nutshell, DTCs are periodically driven systems that manifest a robust response at fractions of the driving frequency. 
To realize such a phase of matter, the system needs to avoid repeated injections of energy from the driving to ``heat-up'' the system, eventually bringing it to an infinite temperature state and to the disruption of the space-time order~\cite{AbaninRigorousPrethermalization}. 
Instead, the response to the external driving has to synchronize with it, stabilizing the emergent phase of matter and blocking the heating; most importantly, the response should survive to generic perturbations and persist in approaching the thermodynamic limit.

Floquet DTCs have emerged as the prototypical theoretical setting~\cite{SachaPRA15_MeanfieldTC,ElsePRL16,KhemaniPRL16,VonKeyserlingkPRB16_AbsoluteStability}. They usually consist in lattice spin models subjected to sudden and periodic pulses of the magnetic field. The goal is to find a suitable strategy to stabilize the temporal order, by suppressing the unbounded heating. Many-body localization (MBL)~\cite{BAA,PhysRevB.75.155111} is a typical setup where spatial disorder provides the mechanism for a strong breakdown of ergodicity. Indeed, in presence of strong disorder, the local excitations generated by the driving protocol are frozen, due to the presence of quasi-local charges~\cite{HuseNandkishore,PhysRevLett.111.127201}, and discrete time-crystals are stabilized \cite{choi2017observation}.

However, MBL is not the only player in the game. A plethora of different mechanisms has been exploblack to realize robust DTC phases (or, at least, transient DTC signatures) in a broad class of generic systems, both theoretically and experimentally. 
They go from pre-thermalization~\cite{PhysRevB.96.094202,AbaninRigorousPrethermalization,ElsePRX17_PrethermalDTC,kyprianidis2021observation,PhysRevX.10.021046,PhysRevLett.120.180603,MachadoPRX20} to emergent Floquet integrability in systems with symmetries~\cite{PhysRevLett.121.093001,PhysRevLett.120.210603,RussomannoPRR20},
from quantum many-body scarring~\cite{turner2018weak,serbyn2021quantum,bluvstein2021controlling} 
to confinement of excitations~\cite{collura2021discrete}.

The search for clean DTCs, namely a system which behaves in a time-crystalline way without the presence of disorder, is indeed the focus of current theoretical investigations. In this perspective, the simplest clean strategy which does allow the existence of a stable DTC relies on the existence of an effective long-range-ordeblack Gibbs ensemble at finite temperature~\cite{AbaninPRB17_EffectiveHamiltonians}. 

Guided by this idea, we explore a natural, yet uncharted, extension of the kicked quantum Ising model in {\em two dimensions}. This provides the quantum counterpart of recent classical results in higher dimensions~\cite{PhysRevLett.127.140602,PhysRevB.104.094308}.

The paper is organized as follows: in Sec.~\ref{sec:Model} we introduce the periodically-kicked transverse-field quantum Ising model. In Sec.~\ref{sec:NumericalEvidences} we discuss numerical evidence of the \textcolor{black}{robustenss of the} discrete time-crystal \textcolor{black}{response} in two-dimensions and we compare the phenomenology with the one-dimensional case. 
In Sec.~\ref{sec:quenchdynamics} we inspect the thermalization of the two-dimensional kicked quantum Ising model in the high-frequency limit. Finally, we draw our conclusions in Sec.~\ref{sec:conclusions}.

\section{Model}\label{sec:Model}
We consider a $d$-dimensional nearest-neighbor quantum Ising model on a hyper-cubic lattice subjected to delta-periodic pulses of the transverse magnetic field. 
The time-dependent Hamiltonian operator of the system reads
\begin{equation}
    \Ham(t) = -J \sum_{\langle jj'\rangle}\Pauli^z_j\Pauli^z_{j'}-\left(\frac{\pi}{2} + \revision{\epsilon}\right) \sum_{n=1}^\infty \delta(t-n\period)\sum_{j}\Pauli^x_j, \label{eq:Ham_kicked_TFI_2d}
\end{equation}
where $J>0$ is the ferromagnetic coupling between nearest-neighbor spins, and $\Pauli^{\alpha}_j$ for $j=1 ,..., N$, with $N=L^d$, $L$ being the size of the hyper-cube, are the standard Pauli matrices. 
Here $\langle jj'\rangle$ denotes the sum over nearest-neighbors. 

In the following, we only consider one- and two-dimensional systems with open boundary conditions (OBC).
The unitary dynamics generated by the time-dependent Hamiltonian in Eq.~\eqref{eq:Ham_kicked_TFI_2d} can be understood as a Floquet dynamics governed by the Ising Hamiltonian evolution operator (we set $\hbar=1$ from now on):
\begin{equation}
    \hat{V} = \ee^{iJ\period \sum_{\me{jj'}} \Pauli^z_j \Pauli^z_{j'} } \;,
\end{equation}
intertwined by sudden imperfect single-spin kicks along the $\hat{x}$-axis
\begin{equation}
    \hat{K}_{\pi/2+\epsilon} = \ee^{i\left(\pi/2+\revision{\epsilon}\right)\sum\limits_j\Pauli^x_j} \;,
\end{equation}
at times $t_n = n\tau = \tau, 2\tau, \dots$, which are integer multiples of the period $\tau$. 
The resulting single-period Floquet operator thus reads 
$\hat{U} = \hat{K}_{\pi/2+\epsilon} \hat{V}$. 

To see the cleanest realization of DTC order, the system is initially prepablack in the fully polarized state with positive magnetization along the $\hat z$ direction, i.e. $\ket{+} = \ket{\uparrow \dots \uparrow }$, where $\ket{\uparrow} (\ket{\downarrow})$ is the eigenvector of the Pauli matrix $\sigma^{z}$ with eigenvalue $+1$ ($-1$). The system experiences a stroboscopic dynamics, and the state after $n$ periods is given by
\begin{equation}
    \ket{\psi_n} = \hat{U}^n \ket{+} = (\hat{K}_{\pi/2+\epsilon} \hat{V})^n \ket{+} \;.
\end{equation}

Since $\hat{K}_{\pi/2+\epsilon} = \hat{K}_{\pi/2}\hat{K}_\epsilon$, and $\hat{K}_{\pi/2} = i^N \hat{P}$ with $\hat{P}=\prod_j \Pauli^x_j$ being the global spin flip operator, such that $\comm{\hat P}{\hat{V}} = 0$,
we can factor out from the evolution the product of all perfect spin flips, thus obtaining
\begin{equation}
    \ket{\psi_n} = \left(\hat{K}_\epsilon \hat{V}\right)^n \hat{K}_{\pi/2}^{n}\ket{+} = 
    (i)^{n}\left(\hat{K}_\epsilon \hat{V}\right)^n \ket{(-)^{n}}.
\end{equation}
This can be interpreted as a stroboscopic change of reference frame, which results in a non-trivial evolution only due to $\left(\hat{K}_\epsilon \hat{V}\right)^n$, on top of perfect alternating jumps between $\ket{+}$ and $\ket{-} = \ket{\downarrow \dots \downarrow}$.

The $\hat{z}$-magnetization after every kick provides information about the persistence of the ferromagnetic order during the stroboscopic dynamics, and 
it is given by 
\begin{align} \label{eq:strobo_mag}
    m(n) = 
    \frac{(-1)^n}{N}&\sum_{j=1}^N\expval{ \big( \hat{V}^{\dagger} \hat{K}_\epsilon^{\dagger} \big)^{n}
    \Pauli^z_j\big(\hat{K}_\epsilon \hat{V}\big)^n}{+} \;. 
\end{align}

Setting $\epsilon = 0$ results in a trivial dynamics in which the system periodically jumps between the two product states $\ket{+}$ and $\ket{-}$, with magnetization being equal to $m(n) = (-1)^n$, thus exhibiting a perfect time-crystal behavior. 
As a matter of fact, one might say that $m(n)$ shows a ``period-doubling'' since $m(n)= m(n+2)$ whereas $\hat{H}(t_n)=\hat{H}(t_{n+1})$.  
Nonetheless, we stress that, in order to realize a stable non-equilibrium DTC phase, the long-range ferromagnetic order has to be robust against arbitrary (sufficiently weak) perturbations in the thermodynamic limit $L\to\infty$.

In order to highlight the emergent stability of such phase in two-dimensions, here we recap some recent results for the one-dimensional case. We refer the reader to \cite{collura2021discrete} for a thorough analysis of the 1D kicked Ising model in absence of disorder with nearest- and beyond nearest-neighbor couplings. 
In particular, in the case of nearest-neighbor interactions in the thermodynamic limit, it is analytically shown, employing free-fermions techniques, and after computing the exact Floquet operator, that $m(n)$ decays exponentially to zero as $\abs{m(n)} \sim \ee^{-\gamma n}$,
where for small perturbations $\epsilon$, the decay rate $\gamma$ scales as $\gamma \sim \abs{\epsilon}^3$.
Moreover, numerical evidence shows that if one considers beyond-next-to-nearest-neighbor interactions in the Hamiltonian (\ref{eq:Ham_kicked_TFI_2d}) up to the $R$-th site, the order-parameter lifetime experiences a qualitative enhancement, and the scaling of the decay rate gets modified into $\gamma \propto \abs{\epsilon}^{2R+1}$. When $R\to\infty$ the system is thus expected to show a long-lasting suppression of the heating induced by the driving. Indeed, the latter may be due to the fact that long-range interacting systems overcome the standard Peierls' argument against the nonexistence of thermal phase transitions in one-dimension~\cite{peierls_1936}. As a consequence, in this framework, we expect that kicked system in two dimensions may support a long-lasting \textcolor{black}{time-crystalline response.}

Moreover, even for large enough kick strength perturbation, which eventually implies an effective paramagnetic stationary state,
we may still expect a long-lived non-equilibrium DTC \textcolor{black}{response} guaranteed by the slow-down of the interface melting in the two-dimensional setup~\cite{balducci2022}.

In the next section, we investigate the robustness against the parameter $\epsilon$ of the DTC \textcolor{black}{response} in the two-dimensional kicked Ising model, via state-of-the-art numerical techniques, and try to infer its thermodynamic behavior.

\section{Numerical results}\label{sec:NumericalEvidences}

\subsection{Exact diagonalization} 
\begin{figure}[t!]
    \centering
    \includegraphics[width=\linewidth]{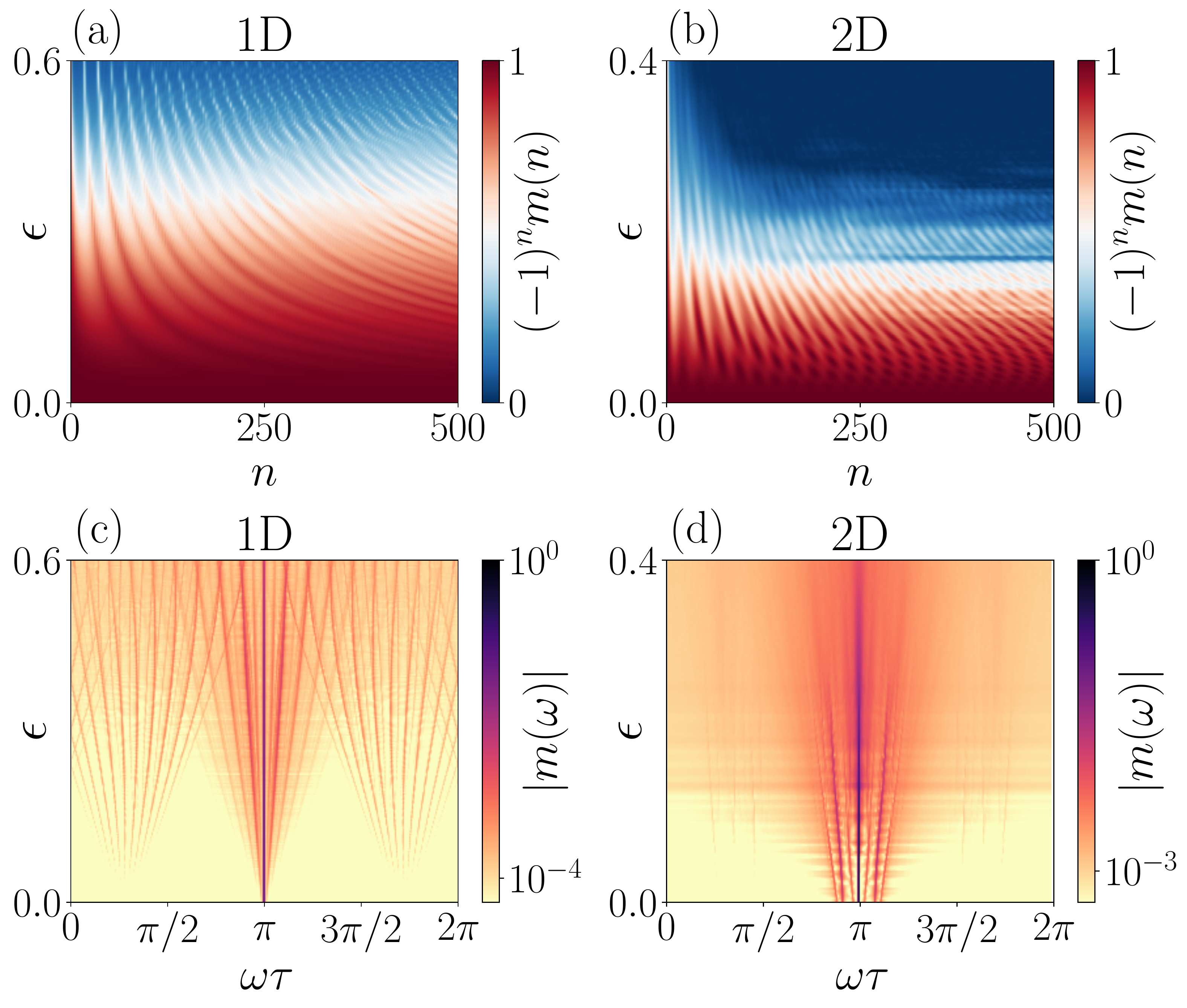}
    \caption{Floquet dynamics of the kicked Ising model with $N=16$ lattice sites and $J\period=1$. (a-b) Stroboscopic evolution of the magnetization in 1D and 2D. (c-d) Absolute value of the discrete Fourier Transform of the corresponding magnetization.}
    \label{fig:ED4x4_OBC}
\end{figure}

We start our analysis by considering small system sizes whose dynamics have been computed via exact diagonalization (ED) techniques~\cite{expm_multiply}. In Fig.~\ref{fig:ED4x4_OBC} we show a color density plot of the stroboscopic evolution of the magnetization and its Fourier transform, for $\epsilon$ ranging in $[0,0.6]$ and fixed energy scale $J\period = 1$. The system consists of $N=16$ lattice sites, arranged in a chain in the 1D case, and in a $4\times 4$ square lattice for the 2D geometry.

The one-dimensional setting is in agreement with the results of \cite{Angelakis_Ising} in which it is shown that finite-size kicked spin chains can sustain a time-crystalline response. However, this behavior is a finite-size effect and does not hold in the thermodynamic limit as shown in \cite{PizziPRB20_CleanDTC,collura2021discrete}. 
Similarly to the 1D case, a very preliminary analysis confirms that in two dimensions does exist a 
region in the $\epsilon-n$ plane wherein stable oscillations of the order parameter are present, at least up to $n=500$ unit periods, and for small system sizes (see Fig.~\ref{fig:ED4x4_OBC}(b)).  
For these values of the parameters, the system synchronizes to the driving and behaves in a time-crystalline way. 
Both in one and two dimensions, by increasing the value of $\epsilon$ we disrupt the spatio-temporal order by letting excitations proliferate in the system thus breaking the DTC response. Interestingly, from the stroboscopic density plots, it seems that the meta-stable ferromagnetic dynamical region extends up to $\epsilon \simeq 0.4$ in the 1D case, while in the 2D case the order starts disappearing at  $\epsilon \simeq 0.15$.
However, while in the 1D setup the system exhibits a smooth transition from one dynamical behavior to the other, in the 2D geometry the transition is remarkably sharper. 
This is confirmed by a thorough analysis of the excitation spectrum via discrete Fourier transform (DFT).

In order to explore the excitation spectrum of the system we plot in Fig.~\ref{fig:ED4x4_OBC}(c-d) the modulus of the discrete Fourier transform $m(\omega)$ of $m(n)$. 
When $\epsilon=0$ there is only the time-crystal characteristic frequency $\omega=\pi$, which corresponds to the period-doubling of the magnetization. By increasing the value of the kick perturbation $\epsilon$, we generate more and more excitations in the system at different frequencies, which will eventually break the order. 
Let us stress that in the 1D case, due to integrability, the excitations that proliferate in the chain result in an extensive set of prominent frequencies (delta-peak in the spectrum) which are present already from $\epsilon = 0^{+}$; they correspond to stable quasi-particle traveling across the system and leading to the melt-down of the DTC behavior at large time. 
Quite interesting, in the 2D geometry, the Fourier spectrum is dominated by only a few peaks for small kick perturbations; as far as the number of such quasi-particle remains finite, we do expect the DTC \textcolor{black}{response to be robust for a finite but long-lasting time in the thermodynamic limit.}

Only for $\epsilon \gtrsim  0.15$ they sharply melt into a continuum of excitations, thus leading to a transition without DTC order.

\begin{figure}[t!]
    \centering
    \includegraphics[width=.85\linewidth]{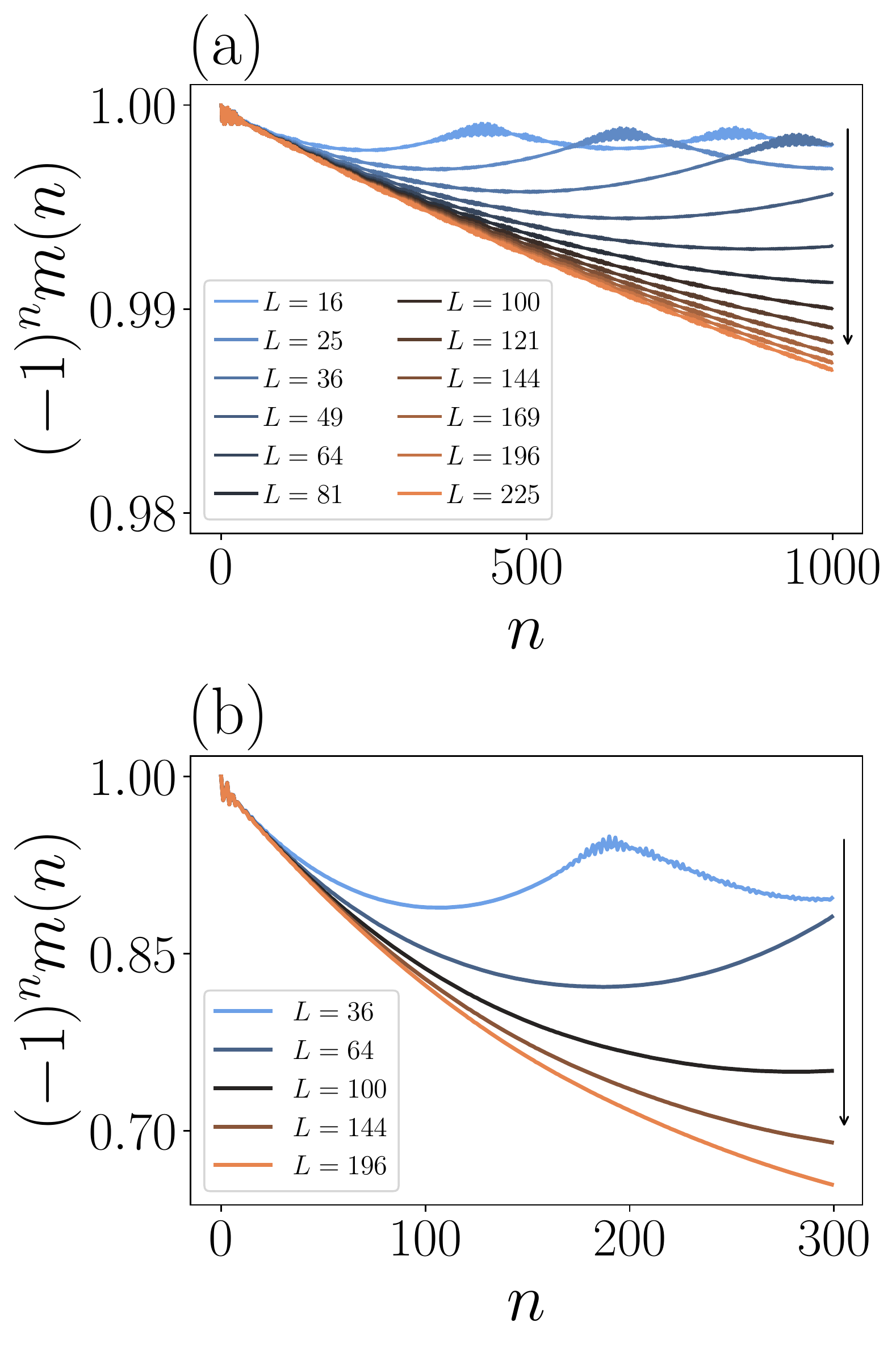}
    \caption{Stroboscopic evolution of the magnetization for the 1D kicked Ising model, with $J\period=1$ and (a) $\epsilon =0.02$, (b) $\epsilon=0.1$. The arrows mark the trend of the stroboscopic magnetization with increasing the system size $L$.}
    \label{fig:1dcomparison}
\end{figure}

\begin{figure*}[t!]
    \centering
    \includegraphics[width=0.85\linewidth]{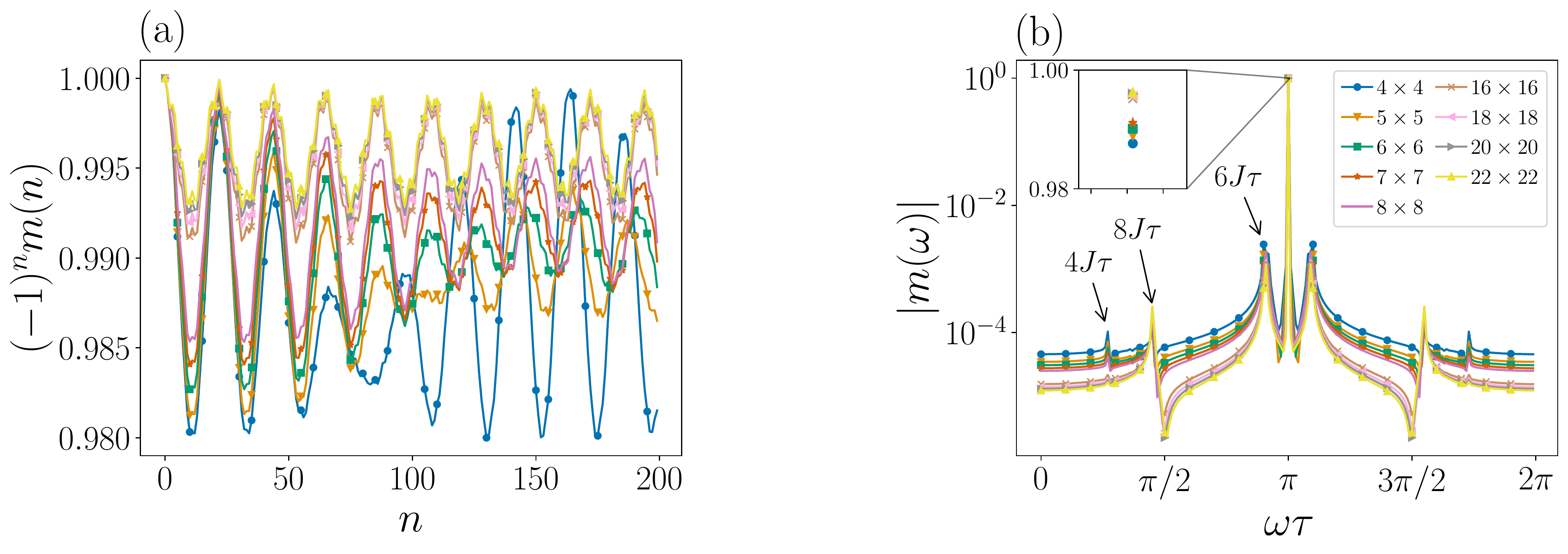}
    \centering
    \includegraphics[width=0.85\linewidth]{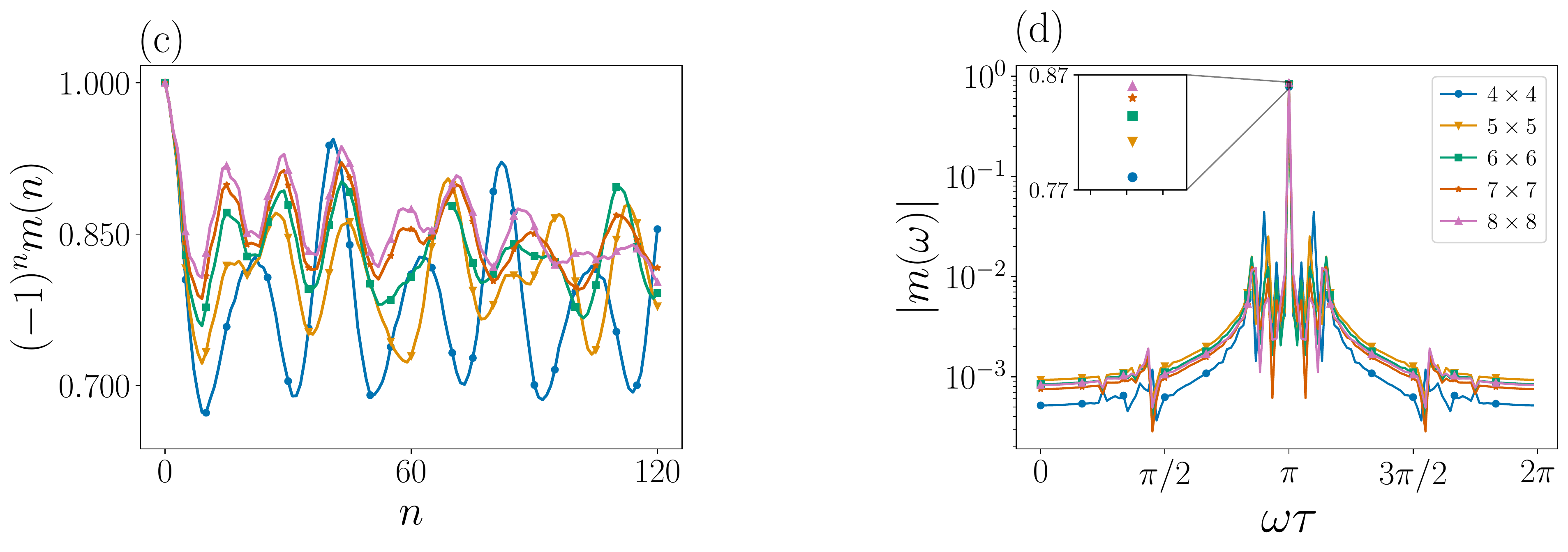}
    \caption{(a,c) Same as in Fig.~\ref{fig:1dcomparison} for the $2D$ case, with $\epsilon=0.02$ in panel (a) and $\epsilon=0.1$ in (c). (b,d) Absolute value of the discrete Fourier transform of the corresponding stroboscopic magnetization. 
    The insets are close-up of the period-doubling peak at $\omega\tau=\pi$. The color code and markers in the legends applies also for the (a,c) panels, respectively. }
    \label{fig:small_epsilon}
\end{figure*}

\subsection{Tensor network time evolution}
The results shown in the previous section are based on the analysis of small systems. 
In order to confirm the fact that in the 2D geometry the kicked Ising model may sustain a DTC phase, we need to rule out whether the evidence we found are artifacts due to finite-size effects.

In particular, we aim at understanding how the recurrences in the order parameter (see Fig.~\ref{fig:ED4x4_OBC}) are possibly caused by finite-size effects rather than being a genuine DTC signature. 
In order to do so, we exploblack the dynamics for larger lattice sizes using Tensor Network (TN) based techniques. Both for 1D and 2D geometries, we used a suitable matrix product state (MPS) representation of the many-body wave function, joined with the correspondent matrix product operator (MPO) representation of the Hamiltonian. 
The non-equilibrium Floquet dynamics has been computed via the time-dependent variational principle (TDVP) algorithm~\cite{PhysRevLett.107.070601,PhysRevB.94.165116,PAECKEL2019167998}.

We start with the one-dimensional case, where we expect that by increasing the system size, the space-time order should disappear.
Indeed, in Figs.~\ref{fig:1dcomparison}(a-b) we show the evolution of the magnetization, with fixed $J\tau = 1$, and $\epsilon = 0.02, 0.1$, respectively. 
We observe that the smaller the system the sooner recurrences appear in the evolution of the magnetization. 
As a matter of fact, with increasing $L$, the curves are approaching the exponentially decaying thermodynamic line. 
As expected, the long-time oscillations are thus finite-size effects whereas in the thermodynamic limit the magnetization does decay to zero.
This will give us a well-grounded numerical reference to compare the forthcoming novel 2D results with.

In Figs.~\ref{fig:small_epsilon}(a-c) we show the same Floquet dynamics for the two-dimensional geometry. 
Remarkably, we find a completely different picture: as the system size is getting larger, the DTC response becomes more robust. In particular, in Fig.~\ref{fig:small_epsilon}(a) where $\epsilon$ is kept small, the many-body wave function remains low entangled, and we are able to simulate fairly large systems, up to $N= 22 \times 22 = 484$ lattice sites, for a relatively large number of kicks. 
This allows us to safely exclude the possibility that finite-size effects may mimic the presence of a stable DTC.  
Indeed, what is crucial here is that, as opposed to the 1D case, passing from the smaller lattice ($4\times 4$) to the larger one ($22\times 22$), we observe an enhancement of the average stroboscopic magnetization, from $\sim 0.985$ to $\sim 0.995$.

When we consider larger values of $\epsilon$, as in Fig.~\ref{fig:small_epsilon}(c), the entanglement that the MPS should encode grows much faster, thus preventing us to consider system sizes bigger than $N=64$ without a sensible numerical error. 
Nonetheless, the qualitative pattern we found is the same as the one illustrated before: increasing the system size stabilizes the space-time order. 
Furthermore, by comparing the 1D case with the 2D case, 
namely, Fig.~\ref{fig:1dcomparison}(b) vs Fig.~\ref{fig:small_epsilon}(c), we observe that, initially, all system sizes manifest the same behavior during the first few kicks. 
However, in 1D, this initial ``transient'' is getting longer by increasing the system size, suggesting that it is not representing a transient at all, but instead the thermodynamic behavior.  
On the contrary, in 2D, the departure from the transient decay starts sooner as the system size is getting larger, showing almost immediately a stable oscillating magnetization. 
As a matter of fact, this suggests that the role of two-dimensional spin-spin interaction is non-trivial and stabilizes the dynamics, as in the case of long-range one-dimensional interactions~\cite{collura2021discrete}. 
We thus expect that the two-dimensional space-time order should persist in the thermodynamic limit, provided a sufficiently small value of the kick perturbation $\epsilon$ is used.

\begin{figure}
    \centering
    \includegraphics[width=\linewidth]{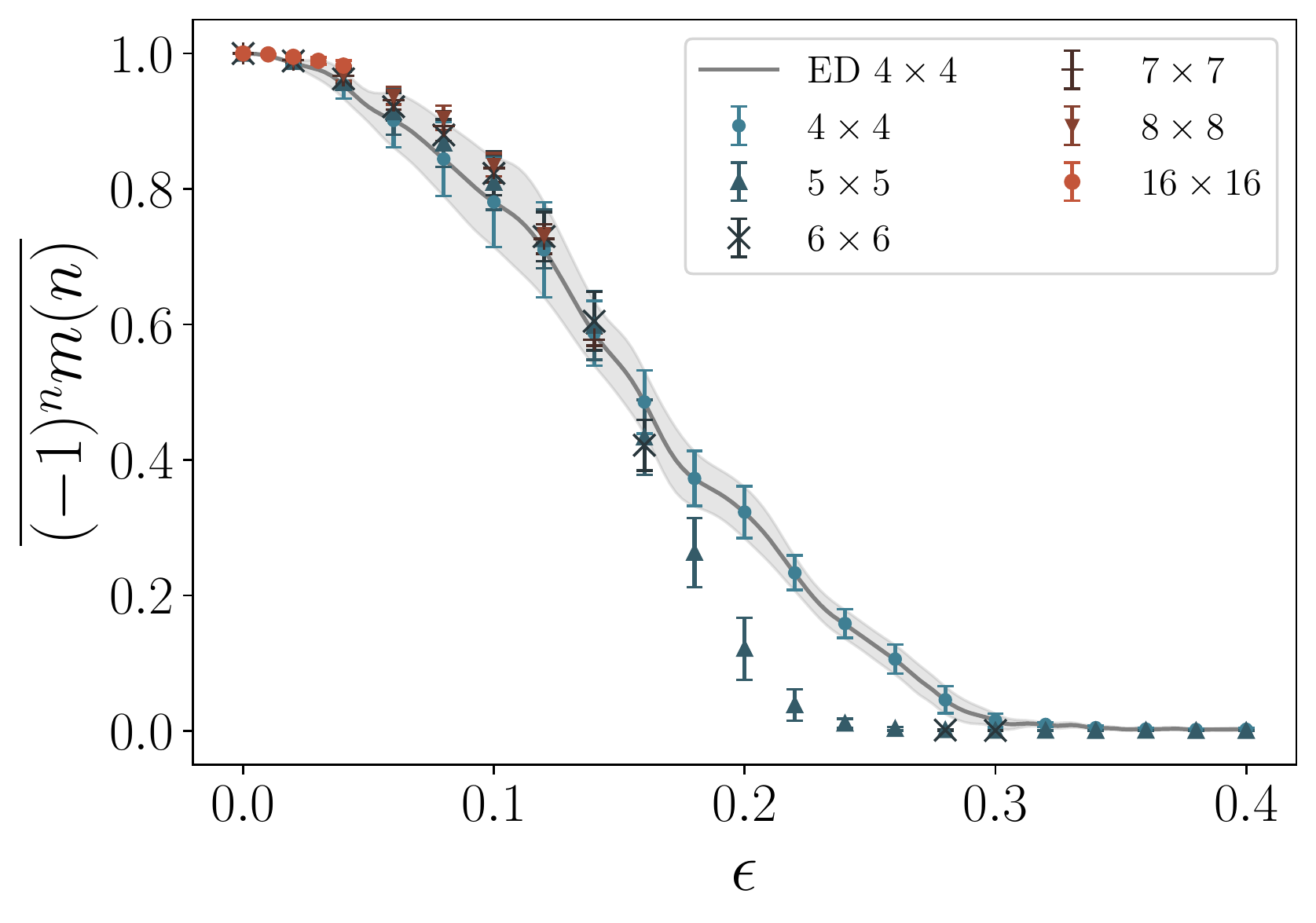}
    \caption{Time-average magnetization vs $\epsilon$ for different system size and $J\tau = 1$. The averages have been take over the time interval $[75\tau,120\tau]$. The solid line represents ED results with shaded region representing the standard deviation. Symbols with error bars are TDVP results.}
    \label{fig:Stationary_J1}
\end{figure}

We further analyze the numerical data by inspecting the power spectrum of the magnetization. 
In Fig.~\ref{fig:small_epsilon}(b) we plot the absolute value of the discrete Fourier transform of the magnetization shown in Fig.~\ref{fig:small_epsilon}(a). 
We marked the peaks of the power spectrum corresponding to the following frequencies
\begin{equation} 
    \omega\period = \pi \pm \tau \Delta E_J \mod 2\pi \;,
\end{equation}
where $\Delta E_J$ measures the energy cost of a single spin-flip on top of the fully polarised state. 
 
In particular, $\Delta E_J=8J$ corresponds to flipping a spin in the bulk of the system, $\Delta E_J=6J$ to flipping a spin in one of the borders of the square lattice, and $\Delta E_J = 4J$, finally, to flipping one of the corner spins. When the driving perturbation $\epsilon$ is weak, single spin-flip excitations represent the lower excited states of the system, and we may expect them to play a dominant role in the dynamics. 
In a $L\times L$ square lattice with OBC we have $4$ excited states with energy gap $\Delta E_J = 4J$ (obtained by a single spin-flip at the corners), $4L$ excited states with gap equal to $\Delta E_J = 6 J$ 
--- the $4(L-2)$ states obtained by a single spin-flip on the $4$ sides, plus $8$ states obtained by flipping two adjacent spins at the corners --- and finally $(L-2)^2$ states with $\Delta E_J = 8J$, obtained via a single spin-flip in the bulk. 

The role of these single-flip excitations in the dynamics reflects in the power spectrum of the magnetization. Apart from the peak at $\omega\tau = \pi$, due to the period-doubling of the magnetization induced by the perfect driving, the next higher contribution comes at $\omega\tau =\pi \pm 6J\tau \mod 2 \pi$, for small $\epsilon$: here the dynamics of the system is mostly confined on lattice boundaries. Notice that the energetically most favorable transitions toward the corner spin-flip states (the $4J\tau$ peak), get suppressed with the system size since their number is not extensive.

Similarly, in Fig.~\ref{fig:small_epsilon}(d) we plot the power spectrum corresponding to the magnetisation reported in Fig.~\ref{fig:small_epsilon}(c). Since here $\epsilon=0.1$ is larger, we are approaching the continuum of the spectrum, and the picture outlined above is going to break down; in practice, the isolated quasi-particle excitations cannot be exactly identified in single spin flips. Nonetheless, the peak at $\omega\tau = \pi$, indicating the presence of DTC order, is getting higher for larger system sizes, meaning that the time-crystalline response is getting more robust.

So far, the numerical finite-size analysis gives evidence of a stable DTC \textcolor{black}{response} for finite values of the kick perturbation $\epsilon$. 
Of course, increasing $\epsilon$ we expect the DTC \textcolor{black}{response} to break down, eventually.
In the following, we characterize, at least qualitatively, such transition. 
In order to do so, we analyzed the long time average of the stroboscopic magnetization $\overline{(-1)^nm(n)}$ as a function of $\epsilon$ for different system sizes, see Fig.~\ref{fig:Stationary_J1}. 
Even though we are far from the thermodynamic limit, we expect a ferromagnetic to paramagnetic dynamical phase transition with
$\epsilon_c \in [0.1,0.2]$. Indeed, for $\epsilon \lesssim 0.15$, the average magnetization data manifest a global increasing trend with the lattice size dimension $L$; on the contrary, when $\epsilon \gtrsim 0.15$, the average magnetization is going to zero as $L$ grows larger.

Notice that, for weak perturbation, $\epsilon<0.05$, we are able to simulate big systems with hundblacks of lattice sites since the auxiliary dimension of the MPS remains relatively small. 
Increasing the strength of the perturbation greatly blackuces the system sizes we can handle, down to $N\approx 36$. For this reason, we cannot quantitatively describe the exact nature of such transition. 

\subsection{Dynamical transition in the relaxation dynamics}\label{sec:OutofEquilibriumDynamics}
To complement the analysis of the dynamical phase transition outlined in the previous section, we further look at the evolution of the stroboscopic magnetization and we study the initial decay of the order parameter before reaching the asymptotic equilibrium. As stated before, in the case of a one-dimensional kicked Ising model it has been proved that in the thermodynamic limit the order parameter decays to zero as $|m(n)|\sim \ee^{-\gamma n}$ with $\gamma \propto |\epsilon|^3$; thus here we inspect whether a similar relation occurs also in the two-dimensional case.

In 2D the situation is more delicate, due to the presence of a reasonable stable DTC \textcolor{black}{response} for $\epsilon \lesssim 0.15$. In this sense, the extrapolated decay only represents a transient toward a stationary value which can be zero or different from zero depending on the non-equilibrium dynamical phase the system ends up.

\begin{figure}
\includegraphics[width=.75\linewidth]{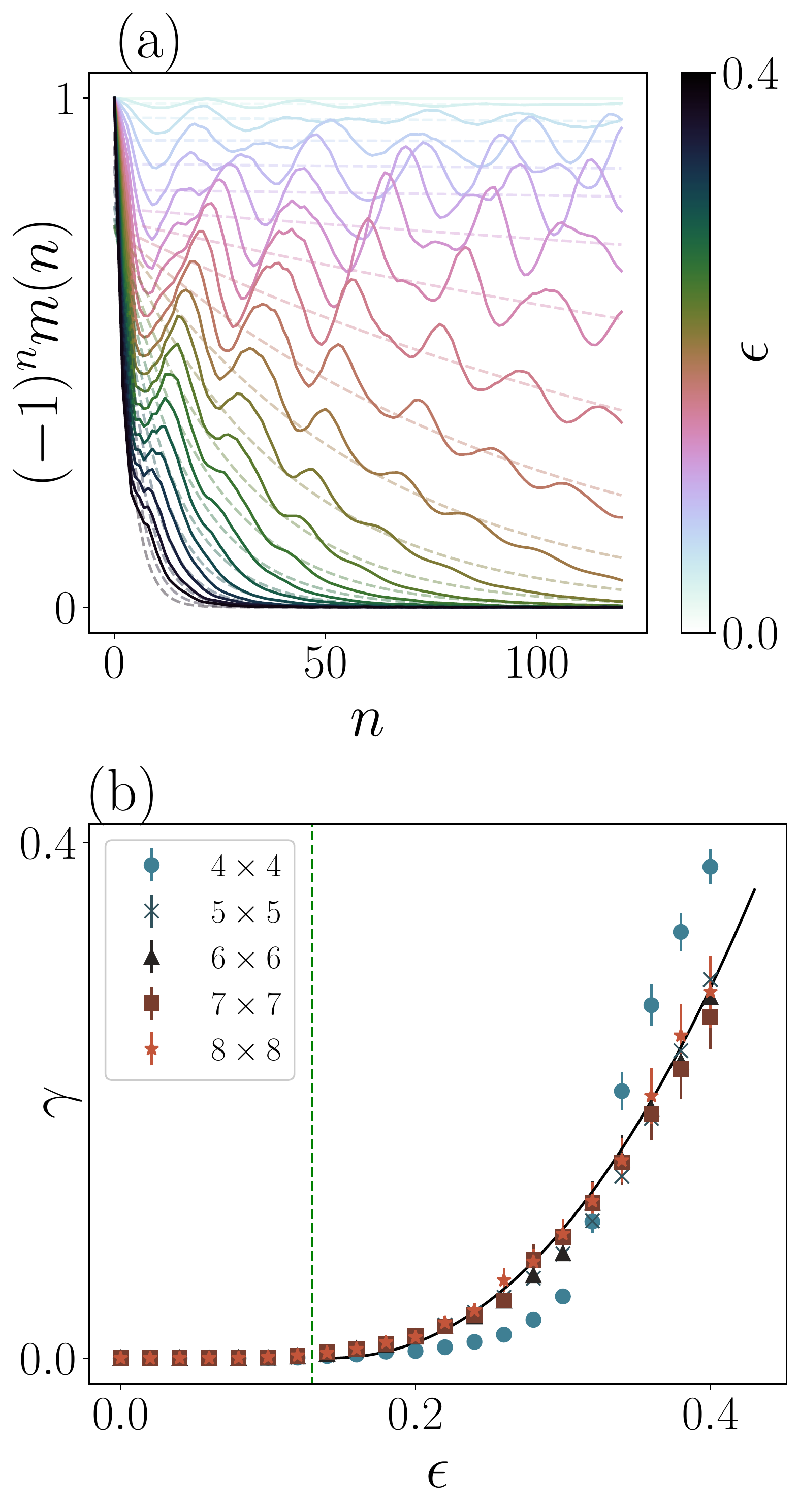}
\caption{(a) Solid lines: evolution of the magnetization for a $5\times5$ system, dashed lines: exponential decay $A\ee^{-n\gamma}$ for fitted values of $\gamma$ and $A$. (b) Decay rates $\gamma$ against $\epsilon$ for different system sizes. In the thermodynamic limit, we expect that the $\gamma$ before the vertical dashed line is identically vanishing.}
\label{fig:relaxation_times}
\end{figure}

In order to obtain an estimate of $\gamma$, we fitted the absolute value of the magnetization with the function $A \ee^{-\gamma n}$, using only the first kicks (whose number depends on how fast the magnetization decays and on the number of kicks we are able to numerically evolve), for the different system sizes. 
In Fig.~\ref{fig:relaxation_times}(a) we show a representative example of the time evolution of the order parameter for a $5\times5$ lattice for different values of $\epsilon$ and their relative best-fit initial transient $\abs{m(n)} \sim  \ee^{-\gamma n}$. We then repeated the procedure for larger system sizes. Once again, we stress that for large systems and high values of $\epsilon$ we are able to time evolve the many-body state just for a few kicks, which are not enough in order to determine the stationary magnetization, but they turn out to be sufficient to evaluate the decay rate. Our analysis suggest that $\gamma$, in the region where the order parameter is expected to decay toward zero (no-DTC \textcolor{black}{response}), increases as a power law \revision{$\gamma \sim |\epsilon|^\alpha$} with $\alpha\approx 4$. 
In practice, by increasing the dimensionality of the problem, the decaying rate of the magnetization is not just quantitatively increased, but rather ``non-perturbatively'' modified, passing from $\gamma \sim|\epsilon|^3$, in 1D, to \revision{$\gamma \sim|\epsilon|^{\alpha}$}, in 2D.
 
This is somehow related to the presence of long-lived domain-wall excitations, which slow down the decay of the fully polarised initial state. 
In practice, when $|\epsilon| \ll J\tau$, the length of the interface between different magnetic domains becomes a quasi-conserved ``charge''. 

\begin{figure*}
\includegraphics[width=\linewidth]{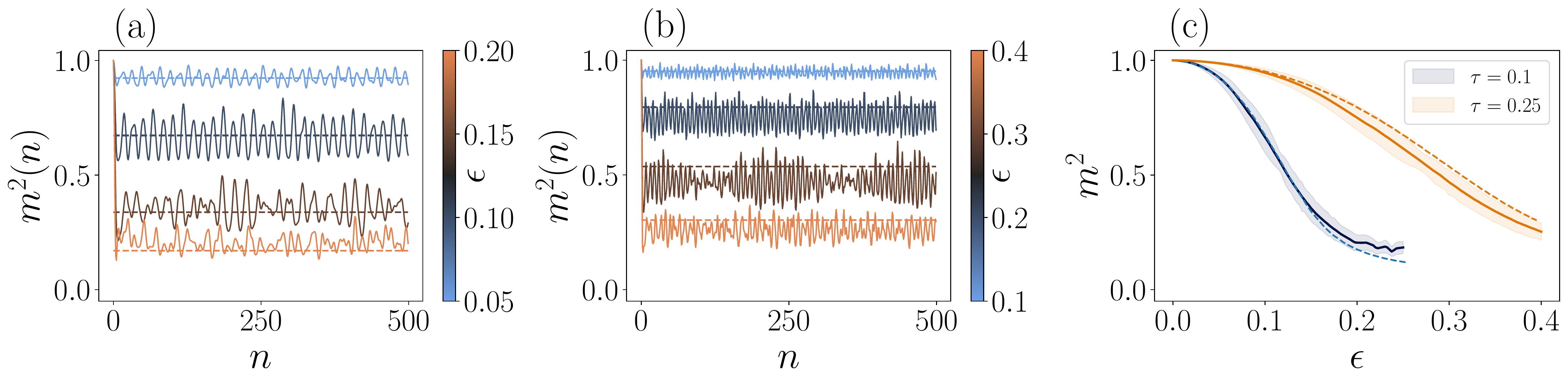}
\caption{(a-b) Stroboscopic evolution of the second moment of the magnetization with $J = 1$ and $\tau=0.1$(a), $\tau=0.25$(b). 
The dashed lines are the relative thermal average computed in the 2D quantum Ising thermal state at late times. (c) Comparison between time averages (solid lines) and thermal averages (dashed lines) as a function of the kick perturbation $\epsilon$. 
The shaded area represents the standard deviation.}
\label{fig:J01Plots}
\end{figure*}

\section{High-frequency limit}\label{sec:quenchdynamics}

The picture emerging from the previous analysis is compatible with a dynamical phase transition driven by the strength $\epsilon$ of the kick perturbation. Besides the evidence of a dynamical transition in the way, the order parameter is decaying --- discussed at the end of the previous section ---, the two phases are also characterized by a change in the stationary properties of the system, which undergoes a transition from \textcolor{black}{a long-lasting} long-range ordeblack ferromagnetic phase to a paramagnetic (disordeblack) phase. 
It is possible to understand the nature of this transition by studying the expected stationary behavior of the system induced by the periodic kicks. As a matter of fact, the system under investigation is non-integrable, thus its stationary properties are expected to be captublack by a canonical Gibbs ensemble, at least in an intermediate ``pre-thermal'' regime ~\cite{AbaninPRB17_EffectiveHamiltonians}, before a final infinite temperature state is possibly attained. 
In order to do so, let us introduce the Floquet Hamiltonian $\Ham_\mathrm{eff}$ as follow
\begin{equation}
    \hat{U} = \hat{K}_\epsilon \hat{V} \equiv \ee^{-i\period\Ham_\mathrm{eff}} \;.
\label{eq:Floquet_Operator}
\end{equation}
If we define $\hat K_{\epsilon}=\ee^{\hat{A}}$ and $\hat V=\ee^{\hat{B}}$, hence $\hat{A} =i\epsilon\sum_j \Pauli^x_j$ and
$\hat{B} =iJ\period\sum_{\me{jj'}}\Pauli^z_j\Pauli^z_{j'}$, the Floquet Hamiltonian can be formally computed from the Baker–Campbell–Hausdorff (BCH) series
\begin{align}
    -i\period \Ham_\mathrm{eff} &= \hat{A}+\hat{B} +\frac{1}{2}\comm{\hat{A}}{\hat{B}}\notag\\&+\frac{1}{12}\left(\comm{\hat{A}}{\comm{\hat{A}}{\hat{B}}}-\comm{\hat{B}}{\comm{\hat{A}}{\hat{B}}}\right) + \cdots \;.
\label{eq:BCHexpansion}
\end{align}
The evolution of the system after $n$ periods, therefore, reads
$\ket{\psi_n} = \ee^{-in \period \Ham_\mathrm{eff}} \ket{+}$, meaning that the evolution is effectively described as the quench dynamics under $\Ham_\mathrm{eff}$ of the generally excited initial state $\ket{\uparrow...\uparrow}$.

Since our system is not integrable  
which implies that Eigenstate Thermalization Hypotesis~\cite{PhysRevA.43.2046,PhysRevE.50.888} is generically expected to hold~\cite{Deutsch_2018,RevModPhys.80.885}, 
we do expect that, after the initial transient, the time averages of local observables should relax toward thermal averages computed in the Gibbs ensemble~\cite{Deutsch_2018,RevModPhys.83.863}
\begin{equation}
    \hat{\rho} = \frac{\ee^{-\beta \Ham_\mathrm{eff}}}{\mathcal{Z}},
\end{equation}
where the partition function is $\mathcal{Z} = \TraceBracket{\ee^{-\beta \Ham_\mathrm{eff}}}$, and the inverse temperature $\beta$ has been fixed by the equivalence between micro-canonical and canonical ensemble, namely
\begin{equation}
    \expval{\Ham_\mathrm{eff}}{+} = \TraceBracket{\hat{\rho}\,\Ham_\mathrm{eff}}\;,\label{eq:def_beta}
\end{equation}
which is nothing more than the conservation of the effective Hamiltonian expectation value. These considerations allow us to compute the thermal average of the order parameter
\begin{equation} \label{eq:thermo_av_mag}
    \me{m}_{\beta} = \frac{1}{N}\sum_j \TraceBracket{\hat{\rho}\, \Pauli^z_j },
\end{equation}
at an effective inverse temperature $\beta$ given by the specific quench protocol, and thus effectively making a bridge between the time-dependent Floquet problem and the finite-temperature behaviour of a system with a very complicated effective Hamiltonian $\Ham_\mathrm{eff}$.

Notice that, this picture holds if and only if the formal series in Eq.~\eqref{eq:BCHexpansion} is convergent. Indeed, the natural guess on the Floquet dynamics is that,
since the driving breaks the energy conservation (where here energy refers to the expectation value of the Ising Hamiltonian), and we are dealing with an ergodic system, we expect to eventually
heat the system towards the infinite-temperature state, i.e., the maximally entropic state. On the other hand, it has been proven that, in the high-frequency driving limit, a quasi-conserved Hamiltonian is expected to constraint the dynamics up to a time $\tau^{*}$ which scales exponentially with the driving frequency; this emergent conservation law prevents the system from heating up and lets it evolve towards a meta-stable long-lived prethermal state~\cite{AbaninPRB17_EffectiveHamiltonians}. 

Since we cannot explore exponentially large times with our numerics, we cannot infer if the plateaux we found in the previous section are actually meta-stable or infinitely lived. Nonetheless, for the non-interacting 1D case, the quadratic Floquet Hamiltonian $\Ham_\mathrm{eff}$ can be computed exactly in the thermodynamics limit, thus implying that the BCH expansion is convergent, and the long-time limit of the Floquet dynamics has to match with the Generalised Gibbs Ensemble constructed with post-quench Bogoliubov fermions~\cite{collura2021discrete}.
In the 2D case, the system is no longer integrable, and the effective Hamiltonian $\Ham_\mathrm{eff}$ is much harder to compute and control: if it is a local Hamiltonian sustaining long-range magnetic order, then a clean DTC \textcolor{black}{response} is expected to survive, at least for a suitably large time, after which the effect of possible small non-local terms might  appear~\cite{AbaninPRB17_EffectiveHamiltonians}.

Again, since the Floquet operator in Eq.~(\ref{eq:Floquet_Operator}) is $\mathbb{Z}_2$-symmetric, the time-crystal \textcolor{black}{response} we observed in the previous section should be related to the ferromagnetic to paramagnetic finite-temperature phase-transition of an effective quantum spin model in two-dimensions. If universality holds, for a sufficiently small perturbation, we may expect that, in the meta-stable regime, all the critical properties of our model can be extracted by those of the two-dimensional transverse field quantum Ising model, whose phase diagram is known since 80s~\cite{PhysRevB.17.1429}. 
The absence of a genuine time-crystal for short-range Hamiltonians in 1D, and its existence in 2D is indeed a consequence of the Peierls argument~\cite{peierls_1936}. 
Indeed, since in a short-range quantum Ising model there is no long-range order at finite temperature in 1D, we expect that in a short-range kicked quantum Ising model there is no DTC order for $\epsilon \neq 0$ in 1D. Analogous considerations hold for the stochastic dynamics of a 2D classical kicked Ising model in which the non-equilibrium stationary state is in the Ising universality class~\cite{Classical_2dDTCIsing}.

We are going to show that this thermodynamic picture holds in a simple case in which we are able to approximate $\Ham_\mathrm{eff}$. 
Indeed, in the high-frequency regime when $J\period,\epsilon \ll 1$ at the lowest order in $J\period\epsilon$ we may neglect all the commutators of $\hat{A}$ and $\hat{B}$ in the BCH expansion and approximate
\begin{equation} \label{eq:efficient2dTFIM}
    \Ham_\mathrm{eff} = -J\sum_{\me{jj'}} \Pauli^z_j\Pauli^z_{j'} 
    - h \sum_j \Pauli^x_j + O\left(J^2\period\epsilon^2\right) \;,
\end{equation}
where $h = \epsilon/\period$ plays the role of an effective transverse field. 
Because of the $\mathbb{Z}_2$-symmetry, the thermal average $\me{m}_{\beta}$ is identically vanishing. In order to study the finite-temperature spontaneous symmetry-breaking we need to evaluate the second moment of the magnetization, namely
\begin{equation}
    \me{m^2}_{\beta} = \frac{1}{N^2} \sum_{jj'}^N \TraceBracket{\hat{\rho}\Pauli^z_j\Pauli^z_{j'}}\;,
\label{eq:thermo_squablack_mag}
\end{equation}
and compare it with the time evolution under the Floquet dynamics
\begin{equation}
    m^2(n) = \frac{1}{N^2}\sum_{jj'}^N \expval{\Pauli^z_j\Pauli^z_{j'}}{\psi_n}\;.
\label{eq:squablack_mag}
\end{equation}

In Fig.~\ref{fig:J01Plots}(a-b) we compare
the asymptotic thermal averages with the stroboscopic dynamics of the magnetisation fluctuations in a $4\times 4$ system, for two different choices of the period $\tau$. 
Thermal averages have been computed with a generalization of the TDVP algorithm after a Wick rotation in imaginary time~\cite{PAECKEL2019167998}.
In particular, we observe that the time-evolved $m^{2}(n)$ is kept oscillating around the corresponding thermal averages; as expected, by  increasing $\epsilon$ the agreement is getting worse, due to the error in the truncation of the BCH expansion. 
Finally, in Fig.~\ref{fig:J01Plots}(c) we compare the thermal equilibrium data of the second moment of the magnetisation with the asymptotic time averages, for two choices of $\tau$. We find a good agreement among the two curves for small enough values of $\epsilon$ and $\tau$, where deviations are typically $O(\max((J\tau)^2,\epsilon^2))$. 
As a matter of fact, this analysis suggests that the stationary properties of the system are well described by the effective 2D Ising Hamiltonian, for a characteristic time which scales as $\sim 1/(\tau J^2\epsilon^2)$.
\\

\section{Conclusions and outlooks}\label{sec:conclusions}

In this paper, we studied the evolution of a clean two-dimensional quantum Ising model periodically kicked with imperfect global spin flips. We compablack the stroboscopic evolution of the magnetization with the one obtained in the one-dimensional kicked Ising model, and by a size-scaling analysis exploiting TDVP calculations, we showed the possibility of realizing a DTC with a two-dimensional clean system. 
Moreover, in the high-frequency limit, we studied a metastable regime wherein local time averages are in perfect agreement with thermal averages computed over an effective Floquet Hamiltonian. 
We pointed out that this quasi-stable time-crystalline response is closely related to the existence of a long-range ordeblack phase at {\it finite} temperature, which may survive for exponentially long times.

Let us remark that the non-equilibrium protocol we have studied can be implemented on currently available quantum platforms, such as trapped ions or superconducting qubits \cite{zhang2017observation,mi2022time,kim2021scalable}. 
In principle, quantum hardware with long coherence times and a small gate noise could outperform the results obtained by means of tensor-network techniques.

Finally, it is worth to further investigate the connection between stable time-crystalline response and finite-temperature long-range order. In this respect, it would be interesting to study periodically-driven dynamics of other different interacting models and lattice topologies, with and without frustration, which may or may not sustain long-range order at finite temperature.

\section*{Acknowledgments} We acknowledge valuable discussions with R. Fazio and E. Tirrito.
G.E.S. was partly supported by EU Horizon 2020 under ERC-ULTRADISS, Grant Agreement No. 834402, 
and his research has been conducted within the framework of the Trieste Institute for Theoretical Quantum Technologies (TQT). 

\section*{Data Availability}
The data that support the plots within this paper and other findings of this study are available from the authors upon request.

\end{document}